\documentclass[floatfix,aps, pra,twocolumn,nobibnotes,groupedaddress,showpacs,nofootinbib]{revtex4-1}
\usepackage{amsfonts}
\usepackage{eurosym}
\usepackage{amsmath, amsthm, amssymb}
\usepackage{tabularx,colortbl}
\usepackage{array}
\usepackage{amscd}
\usepackage{dsfont}
\usepackage{graphicx}
\usepackage{color}

\newcommand{\bra}[1]{\langle #1|}
\newcommand{\ket}[1]{|#1\rangle}

\begin{document}

\title{Topology of Entanglement Evolution of Two Qubits}
\author{Dong Zhou, Gia-Wei Chern, Jianjia Fei, and Robert Joynt}
\date{\today}
\affiliation{Physics Department, University of Wisconsin-Madison, Madison, Wisconsin
53706, USA}

\begin{abstract}
The dynamics of a two-qubit system is considered with the aim of a general
categorization of the different ways in which entanglement can disappear in
the course of the evolution, {e.g., entanglement sudden death}. 
The dynamics
is described by the function $\vec{n}(t)$, where $\vec{n}$ is the $15$-dimensional 
polarization vector. This representation is particularly useful
because the components of $\vec{n}$ are direct physical observables, there
is a meaningful notion of orthogonality, and the concurrence $C$ can be
computed for any point in the space.  We analyze the
topology of the space $S$ of separable states (those having $C=0$), and the
often lower-dimensional linear dynamical subspace $D$ that is characteristic
of a specific physical model. This allows us to give a rigorous
characterization of the four possible kinds of entanglement evolution. Which
evolution is realized depends on the dimensionality of $D$ and of $D\cap S$,
the position of the asymptotic point of the evolution, and whether or not the
evolution is ``distance-Markovian", a notion we define. 
We give several examples to illustrate the general
principles, and to give a method to compute critical points.
We construct a model that shows all four behaviors.
\end{abstract}

\pacs{02.40.Pc,03.65.Ud,03.65.Yz,03.67.Mn}
\maketitle

\section{Introduction}

Entanglement is one of the most intriguing aspects of quantum physics and is
known to be a useful resource for quantum computation and communication \cite%
{Horodecki09,Nielsen00Chuang}. However, its structure and evolution in
time are not fully understood even for simple systems such as two qubits
where a relatively computable entanglement measure, the Wootter's
concurrence $C$, is available \cite{Hill97,*Wootters98,*Wootters01}. 
The difficulty resides in
the high dimensionality of state spaces (fifteen for two qubits) and the
non-analyticity of the definition of the entanglement measure $C$.

In the presence of external noise, pure states become mixed and entanglement
degrades. These are distinct but related issues. The completely mixed state
(density matrix proportional to the unit matrix) is separable: it has $C=0$.
Pure states, on the other hand, can have any value of $C$ 
    between $0$ and $1$ inclusive. 
Purity (measured,
for example, by the von Neumann entropy) tends to decrease monotonically and
smoothly with time under Markovian evolution.  The
same is not true for entanglement. Apart from the expected smooth half-life
(HL) decaying behavior, the sudden disappearance of entanglement has been
theoretically predicted and experimentally observed \cite
{Yu09review,Almeida07,*Laurat07,*Xu10}. Widely known as entanglement
sudden death (ESD), this non-analytic behavior has been shown to be a
generic feature of multipartite quantum systems regardless of whether the
environment is modeled as quantum or classical \cite
{Zyczkowski01,Yu04PRL,Yu06PRL,*Yu06Opt,*Yu10Opt,
Huang07,Cao08,Al-Qasimi08,Hernandez08,Roszak10,Al-Amri09,Ikram07,Vaglica07,
Leon09,Bellomo07,*Bellomo08,Zhou10QIP}. While the monotonic decrease of $
C(t) $ is usually associated with Markovian evolution, non-Markovian
evolution can also lead to entanglement sudden birth (ESB). It is believed
to be related to the memory effect of the (non-Markovian) environment \cite
{Ficek06,*Mazzola09,
*Lopez08,Zyczkowski01,Yonac06,*Yonac08,Bellomo07,*Bellomo08,Zhou10QIP}.
Although most investigations have been focused on two-qubit systems and we
will also focus on this case, ESD and ESB have been shown to exist in
multi-qubit systems, and even in quantum systems with infinite dimensional
Hilbert spaces, such as harmonic oscillators \cite{Ann07,
Weinstein09jan,*Weinstein09may,
*Weinstein09aug,Aolita08,Lai08,Paz08,*Paz09, Lin08,Vasile09}.

Our aim in this paper is to formulate the problem of the evolution of
entanglement in two-qubit systems in the polarization vector space, and to
show that this formulation leads naturally to a categorization of
entanglement evolutions into four distinct types, generalizing and making precise the
concepts of HL, ESD, and ESB behaviors. It turns out that these categories
are consequences of certain topological characteristics of a model. To show
this, we proceed as follows. Sec. \ref{sec:vec} characterizes in detail two
manifolds in the polarization vector space: the manifold of admissible
physical states and the manifold of separable states. Sec. \ref{sec:dyn}
presents the concept of a dynamical subspace - a manifold that is associated
with a physical model, and then gives several concrete examples of models of
increasing complexity. In Sec. \ref{sec:dyn} we also show how to compute 
critical points: parameter values that separate one behavior from another. 
In Sec. \ref{sec:Ecat}, we prove that our categorization scheme is exhaustive.
Sec. \ref{sec:conc} presents the final results and discussion.

\section{Entanglement in Polarization Vector Space}

\label{sec:vec}

The universality of the various entanglement behaviors suggests that they
are derived from some structural property of entanglement in the physical
state space, and that the system dynamics can be viewed as a probe of that
property. To state this property precisely, we need to first characterize
the space of all admissible density matrices $\rho $, or equivalently, the
space of all admissible polarization vectors.

For two qubits, the polarization vector $\vec{n}=(n_{1},n_{2},\ldots
,n_{15}) $ is defined by the equation 
\begin{equation}
\rho =\frac{1}{4}I_{4}+\frac{1}{4}\sum_{i=1}^{15}n_{i}\mu _{i},
\label{eq:polarVec}
\end{equation}
where $I_{4}$ is the $4\times 4$ unit matrix and the $\mu _{i}$ are the
generators of $SU(4)$, satisfying 
\begin{equation}
\mu _{i}=\mu _{i}^{\dagger },\quad \text{Tr~}\mu _{i}=0,\quad \text{Tr~}\mu
_{i}\mu _{j}=4\delta _{ij}.
\label{eq:su4}
\end{equation}
For our purposes the $\mu _{i}$ are most conveniently chosen as 
\begin{equation}
\mu _{\alpha \beta }={\sigma }_{\alpha }\otimes {\sigma }_{\beta },
\label{eq:pauli}
\end{equation}
where $\sigma _{\alpha }$ acts on the first qubit and $\sigma _{\beta }$
acts on the second qubit. $\alpha $ and $\beta $ sum over the $2\times 2$
unit matrix $I$ and the Pauli matrices $X$, $Y$ and $Z$. Thus, in Eqs. \ref{eq:polarVec} and \ref{eq:su4}, $i$ is regarded as a composite index of $%
\alpha $ and $\beta $, but the $\sigma _{\alpha }=\sigma _{\beta }=I$ term
is singled out. This space has the usual Euclidean inner product (which
corresponds to the Hilbert-Schmidt inner product on the the density
matrices), and the inner product induces a metric and a topology in the
usual fashion. The components of $\vec{n}$ are physical observables and can
be calculated by $n_{i}=\text{Tr~}\rho \mu _{i}$. For example, the average
value of the z-component of the spin of the first qubit is $\left\langle
Z\otimes I\right\rangle =\text{Tr~}[\rho (Z\otimes I)]=n_{ZI}$.  The six
components $n_{IX}$, $n_{IY}$, $n_{IZ}$, $n_{XI}$, $n_{YI}$ and $n_{ZI}$
represent physical polarizations of spin qubits. The other nine components ($%
n_{XX}$, etc.) are inter-qubit correlation functions. The most common name
for $\vec{n}$ is \textquotedblleft polarization vector", but
\textquotedblleft coherence vector" and \textquotedblleft generalized Bloch
vector" are also in use. We note that different normalizations in Eqs. \ref%
{eq:polarVec} and \ref{eq:su4} are used in the literature \cite%
{Kimura03,*Byrd03,Mahler98,Alicki_Lendi,Joynt_2009,Byrd11}.

The generators $\mu_i$ satisfy 
\begin{align}
\mu_i\mu_j = \delta_{ij} I + (if_{ijk}+d_{ijk})\mu_k
\end{align}
where $f_{ijk}$ is totally anti-symmetric and $d_{ijk}$ is totally symmetric.
These structure constants can be found in Appendix \ref{sec:section}.

Eq. \ref{eq:polarVec} holds for a $4$-level system. It has an obvious
generalization to $N$-level systems; the $\mu _{i}$ just become the $N^{2}-1$
generators of $SU(N)$. For $N=2$, $\vec{n}$ is the usual Bloch vector in a
real $3$-dimensional vector space. It is important to stress that the
correspondence between $\rho $ and $\vec{n}$ is one-to-one; they give
completely equivalent descriptions of the physical system. Certain physical
concepts have geometric interpretations when stated in terms of $\vec{n}$;
as we shall see below. This is not so true of $\rho $. In our opinion,
$\vec{n}$ is the more convenient quantity for most purposes. 
$\rho$ has been the traditional language in which to describe mixed states,
but some experimental groups now favor $\vec{n}$ \cite{Chow10,DiCarlo10}.

We shall refer to the set of all admissible $\vec{n}$ as $M$, the state
space.  What shape does $M$ have? Eq. \ref{eq:polarVec} guarantees that $%
\rho $ is Hermitian and has unit trace. To guarantee that $\rho $ is
positive (all its eigenvalues are non-negative), 
we also need the condition that all coefficients $a_{i}$ of the
characteristic polynomial $\text{det}(xI_{N}-\rho)=\sum_{j=0}^N(-1)^ja_jx^{N-j}$
are non-negative \cite {Horn85}. 
    Note $a_0=1$ by definition.
 
For two-qubit systems there are four of them, which are 
\begin{align}
1!a_{1}& =\text{Tr~}\rho =1, \\
2!a_{2}& =1-\text{Tr~}\rho ^{2}, \\
3!a_{3}& =1-3\text{Tr~}\rho ^{2}+2\text{Tr~}\rho ^{3}, \\
4!a_{4}& =1-6\text{Tr~}\rho ^{2}+8\text{Tr~}\rho ^{3}+3\left( \text{Tr~}\rho
^{2}\right) ^{2}-6\text{Tr~}\rho ^{4}.
\end{align}
Note that $a_{1}\geq 0$ is trivially satisfied for all density matrices.

For one qubit $N=2$, $\vec{n}$ is the usual Bloch vector, and only the $%
a_{2}\geq 0$ constraint applies. Thus the positivity requirement is that $%
\left\vert \vec{n}\right\vert \leq 1$ and $M$ is the familiar $3$%
-dimensional spherical volume. For the $2$-qubit $N=4$ case that we are
concerned with, there are cubic and quadratic inequalities to be satisfied,
so the surface that bounds $M$ is not so simple. The main point, however, is
that $M$ \textit{is convex}: the line joining any two points in $M$ is also
in $M$. This follows from the convexity argument for $\rho $: if $\rho
_{1}$ and $\rho _{2}$ are positive, then so is $s\rho _{1}+(1-s)\rho _{2}$
for all $0\leq s\leq 1$. This argument clearly also holds for $\vec{n}$. 

All of the positivity requirements can be written in terms of $\vec{n},$ but
the higher-order ones are fairly complicated.  The requirement $1-$Tr 
$\rho ^{2}\geq 0$ is of particular interest, since it has a simple
expression in terms of $\vec{n}:$ 
\begin{eqnarray*}
0 &\leq &1-\text{Tr}\left( \rho ^{2}\right) \\
&=&1-\frac{1}{16}\left[ \text{Tr }I_{4}+2\sum_{i}^{15}n_{i}\text{Tr}\mu
_{i}+\sum_{i,j=1}^{15}\ n_{i}n_{j}\text{Tr}\mu_{i}\mu_{j}\right] \\
&=&\frac{3}{4}-\frac{1}{4}\left\vert \vec{n}\right\vert ^{2},\text{ or ~} \left\vert \vec{n}\right\vert ^{2} \leq 3.
\end{eqnarray*}%
Hence the vectors in $M$ lie within a sphere of radius $\sqrt{3}$. 
Technically, $M$ is a $15$-dimensional manifold with boundary. 
We will follow physics
usage and also employ the term ``space" for $M$, though of course it is not
closed under vector addition. Note that pure states satisfy Tr $\rho ^{2}=1$,
 so the pure states are a subset of the $14$-sphere
in $M$ with $|\vec{n}|=\sqrt{3}$.
    To be more specific, the two-qubit pure states $\ket{\psi}\bra{\psi}$ are of measure zero on that
    sphere since they can be parametrized by $6$ real parameters 
    \begin{align*}
    \ket{\psi} =& \cos{\theta_1}\ket{00}+e^{i\phi_1}\sin{\theta_1}\sin\theta_2 
                            \ket{01} \\
        &+ e^{i\phi_2}\sin\theta_1\cos\theta_2\cos\theta_3\ket{10}\\
        &+ e^{i\phi_3}\sin\theta_1\cos\theta_2\sin\theta_3\ket{11}.
    \end{align*}
    An overall phase has been dropped in writing $\ket\psi$
    since it does not appear in $\rho$.
 
\begin{figure}[tbp]
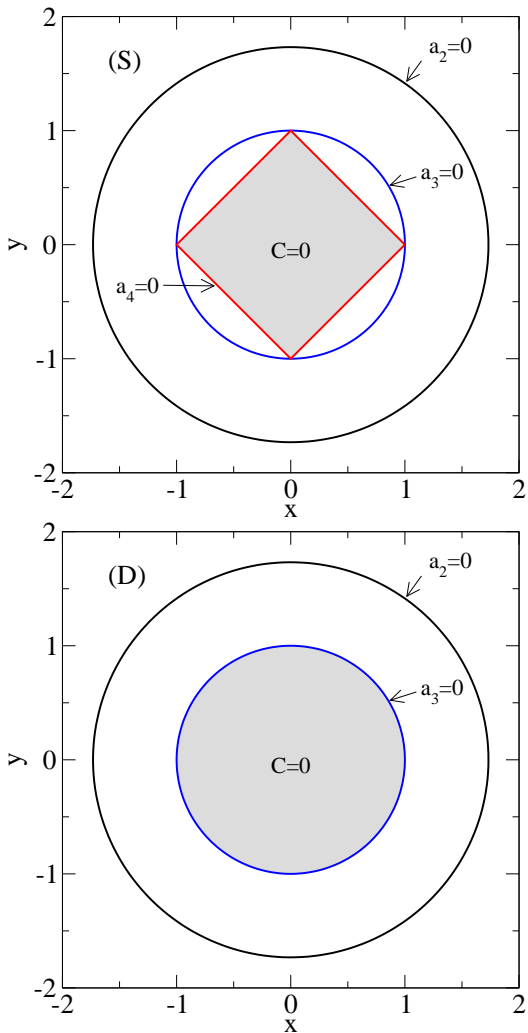

\centering\includegraphics*[width=.8\linewidth]{sec_A.eps} \centering%
\includegraphics*[width=.8\linewidth]{sec_B.eps}
\caption{(Color online) Two dimensional sections of the full two-qubit state
space $M$, keeping only two components of the polarization vector non-zero.
Only two inequivalent sections are possible: the disc section (D) and 
square section (S). The physical states are colored grey. $a_2=0$ is
outlined as black, $a_3=0$ blue and $a_4=0$ red. All physical states in
these two dimensional section are separable states. }
\label{fig:2d}
\end{figure}

Further insight into the shape of $M$ can be gained by noting that $M$ must
be invariant under local unitary transformations (rotations of one spin at a
time), which means that $M$ has cylindrical symmetry around the $6$
single-qubit axes. This is verified by making some $2$-dimensional sections
of $M$ with exactly two components of $\vec{n}$ non-zero. In contrast to
Ref. \cite{Jakobczyk01} where a different basis was used 
(generalized Gell-Mann matrices), we find only two
types of shapes, as shown in Fig. \ref{fig:2d} and tabulated in Table \ref{tab:2d}. 
Using the structure constants $d_{ijk}$ and $f_{ijk}$, it can be shown that 
the square and disc sections are the only possibilities along
the $n_i-n_j$ plane. 
    When $\mu_i$ commutes with $\mu_j$ the section is a square and $\mu_i$ 
    anticommutes with $\mu_j$ the section is a circular disc.
Details can be found in Appendix \ref{sec:section}.

\begin{table}[tb]
\centering
\begin{tabular}{cc|cccccccccccccccc}
\hline
&  & 1 & 2 & 3 & 4 & 5 & 6 & 7 & 8 & 9 & 10 & 11 & 12 & 13 & 14 & 15 &  \\ 
\hline\hline
IX & 1 &  &  &  &  &  &  &  &  &  &  &  &  &  &  &  &  \\ 
IY & 2 & D &  &  &  &  &  &  &  &  &  &  &  &  &  &  &  \\ 
IZ & 3 & D & D &  &  &  &  &  &  &  &  &  &  &  &  &  &  \\ 
XI & 4 & S & S & S &  &  &  &  &  &  &  &  &  &  &  &  &  \\ 
XX & 5 & S & D & D & S &  &  &  &  &  &  &  &  &  &  &  &  \\ 
XY & 6 & D & S & D & S & D &  &  &  &  &  &  &  &  &  &  &  \\ 
XZ & 7 & D & D & S & S & D & D &  &  &  &  &  &  &  &  &  &  \\ 
YI & 8 & S & S & S & D & D & D & D &  &  &  &  &  &  &  &  &  \\ 
YX & 9 & S & D & D & D & D & S & S & S &  &  &  &  &  &  &  &  \\ 
YY & 10 & D & S & D & D & S & D & S & S & D &  &  &  &  &  &  &  \\ 
YZ & 11 & D & D & S & D & S & S & D & S & D & D &  &  &  &  &  &  \\ 
ZI & 12 & S & S & S & D & D & D & D & D & D & D & D &  &  &  &  &  \\ 
ZX & 13 & S & D & D & D & D & S & S & D & D & S & S & S &  &  &  &  \\ 
ZY & 14 & D & S & D & D & S & D & S & D & S & D & S & S & D &  &  &  \\ 
ZZ & 15 & D & D & S & D & S & S & D & D & S & S & S & S & D & D &  &  \\ 
\hline
\end{tabular}%
\caption{Two dimensional sections of the full two-qubit state space $M$,
keeping only two components of the polarization vector non-zero. Type S is
the square section, while type D is the disc section. The table is symmetric
thus the upper part is omitted. The diagonal entries do not correspond to
two dimensional sections. }
\label{tab:2d}
\end{table}

The discs correspond to the local rotations between
single-qubit-type axes, such as the $n_{IX}-n_{IY}$ section. If, on the
other hand we rotate from a definite polarization state of qubit 1 to a
definite polarization state of qubit 2, we find a square cross-section;
examples are the $n_{IX}-n_{XI}$ or $n_{IX}-n_{YI}$ sections. Rotations of $%
\vec{n}$ that mix single--qubit-type and correlation-type directions can be
of either shape; the $n_{IX}-n_{XX}$ section is square, while the $%
n_{IX}-n_{YY}$ section is a disc. Finally, rotations between
correlation-type directions can have either shape. The $n_{XX}-n_{XY}$
section corresponds to a local rotation of qubit 2; hence it is a disc.
Rotations involving both qubits, such as that which generates the $%
n_{XX}-n_{YY}$ section, generally give square sections.

We may conclude that $M$ is a highly dimpled ball, perhaps most similar in
shape to a golf ball. Its minimum radius is $\left\vert \vec{n}\right\vert
=1 $ and its maximum radius is $\left\vert \vec{n}\right\vert =\sqrt{3}$.

Since our aim is to quantify entanglement in $M$, we need an entanglement
measure. We will employ $C$, the concurrence of Wootters \cite{Wootters98}. 
The concurrence varies from $0$ for separable states to $1$ for maximally
entangled state, i.e., the Bell-like states. It is defined as $C=\max \{0,q\}$%
, and 
\begin{equation*}
q={\lambda _{1}}-{\lambda _{2}}-{\lambda _{3}}-{\lambda _{4}}~,
\end{equation*}%
where $\lambda _{i}$ are the square roots of the eigenvalues of the matrix $%
\rho _{AB}\tilde{\rho}_{AB}$ arranged in decreasing order and 
\begin{equation}
\tilde{\rho}_{AB}=(\sigma _{y}^{A}\otimes \sigma _{y}^{B})\rho _{AB}^{\ast
}(\sigma _{y}^{A}\otimes \sigma _{y}^{B}),
\label{eq:flip}
\end{equation}
is a spin-flipped density matrix. $\rho _{AB}^{\ast }$ is the complex
conjugate of the density matrix $\rho _{AB}$. It is not possible to write
the function $C\left( \vec{n}\right) $ in a simple explicit form unless
further restrictions on $\vec{n}$ apply \cite{Rau09}, but it is clear from
the form of the continuous function $q\left( \vec{n}\right) $ and the
presence of the $\max $ function that $C$ is a continuous but not an
analytic function of $\vec{n}$.

We next consider $S$, the manifold of separable states, which we define as
those $\vec{n}$ for which the concurrence vanishes: $C(\vec{n}) =0$. 
$S$ is a subset of $M$ and $M\backslash S$ is the set of entangled
states. $S$ includes the origin since $q(\vec{0}) =-1/2$ and $ C(\vec{0}) =\max 
\{ 0,q(\vec{0})\} =0$. 
Since $q(\vec{n})$ is continuous, $S$ actually includes
a ball of finite radius about the origin: it can be shown that if $
\left\vert \vec{n}\right\vert \leq 1/\sqrt{3}$, then $\vec{n}\in S$ \cite%
{Zyczkowski98}. Thus the manifold of separable states has finite volume in $M
$: $S$ is also $15$-dimensional.  We will also refer below to the interiors
and boundaries of $M$ and $S$ and denote these by Int$(M)$, $B_M$, Int$(S)$, 
and $B_S$.  Since the various sets we encounter in this paper are not linear 
subspaces, we need the general topological definitions of ``boundary", ``interior" and
``dimension".  These may be found, for example, in Ref. \cite{Munkres74}.

$S$ is also a convex set. What else can we say about the shape of $S$? It is easily seen that the
surface of $S$, like that of $M$, is rather non-spherical. Indeed $C=0$
along any of the basis vector directions. Coupled with the fact that $S$ is
convex, we see that $S$ must contain a large hyperpolygon with vertices at $%
\vec{n}=\left( \pm 1,0,0,...,0\right)$, $\vec{n}=\left( 0,\pm
1,0,0,...,0\right)$, etc. $C\left( \vec{n}\right) $ is invariant under
local rotations, so it has the same hyper-cylindrical symmetry as $M$. 
{Again we may consider $2$-dimensional sections in order to understand the shape of the surface.} 
Two examples are shown in Fig. \ref{fig:2d}.

A simple-sampling Monte Carlo study shows that there are more entangled states than 
separable states in $M$.
The details can be found in Appendix \ref{sec:vol}.

\section{dynamical evolution in $S$}

\label{sec:dyn}

\subsection{introduction}

\begin{figure}[tbp]
\centering\includegraphics*[width=.8\linewidth]{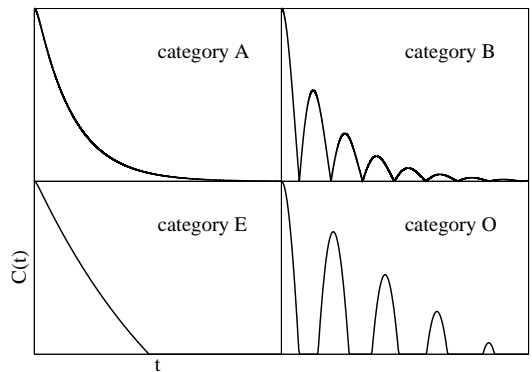}
\caption{Four categories of entanglement evolution. $\mathcal{A}$:
approaching. $\mathcal{B}$: bouncing. $\mathcal{E}$: entering. $\mathcal{O}$%
: oscillating. }
\label{fig:cat}
\end{figure}

The dynamical evolution or trajectory of a quantum system is a function 
$\vec{n}\left( t\right) $ with $t\in [0,\infty )$ and $\vec{n}\in M$.
 The initial point is $\vec{n}\left( 0\right) $ and, in the cases of
interest here, the trajectory approaches a limiting point as $t\rightarrow
\infty $ and we can define $\vec{n}_{\infty }=\lim_{t\rightarrow \infty
}\vec{n}(t)$. The entanglement evolution is the associated function 
$C\left( t\right) =C\left( \vec{n}\left( t\right) \right)$. $C\in [0,1]$.  
For studies of decoherence the main interest is in entanglement
evolutions such that $C(0)>0$ and $C(\infty )=0$, i.e., the system starts 
in an entangled state and ends in a separable state.
Four distinct categories of entanglement evolution of this type have been seen 
in model studies \cite{Zhou11QIP}. 
They are shown in Fig. \ref{fig:cat}. 
These four categories are topologically distinct, as may be seen by considering the 
set $T_0\equiv\{t|C(t)=0\}$. In category $\mathcal A$, $T_0$ is the null set; in
category $\mathcal E$, $T_0$ is a single infinite interval; in category $\mathcal B$,
$T_0$ is a set of discrete points; in category $\mathcal O$, $T_0$ is a union of 
disjoint intervals.

{These categories also reflect }
how the trajectory traverses $S$ and $M\backslash S$. Entanglement
evolutions in category $\mathcal{A}$ approach the boundary of separable and
entangled regions asymptotically from the entangled side. The trajectories
never hit $S$ while the decrease in entanglement may or may not be
monotonic, as seen in Ref. \cite{De11}. Entanglement evolutions in
category $\mathcal{B}$ bounce off the surface of $S$ at finite times but
never enter $S$. Overall, entanglement diminishes nonmonotonically.
Entanglement evolutions in category $\mathcal{E}$ enter $S$ at finite time
and entanglement stays zero afterwards. This is the typical ESD behavior.
Entanglement evolutions in category $\mathcal{O}$ give ESB: after ESD,
entanglement suddenly appears after some dark period.

We shall focus on models with associated linear maps $\Lambda$, i.e., 
$\rho(t) = \Lambda(t)[\rho(0)]$. More general
non-linear models may be contemplated, but they seem to have unphysical
features \cite{Preskill}. 
It is known that $\Lambda(t)$ is 
completely positive (CP) if and only if there exists a set of operators
$\{E_a\}$ such that \cite{Nielsen00Chuang}
\begin{align}
\Lambda[\rho(0)] = \sum_a E_a \rho(0) E_a^\dagger. \label{eq:Lambda}
\end{align}
We require $\Lambda$ to be trace preserving so that it maps density
matrix to density matrix. This condition is equivalent to the completeness 
condition $\sum_a E_aE_a^\dagger=I$.
In terms of the polarization vector, the dynamics is 
described by an affine map $\Upsilon(t)$ acting on the initial
polarization vector $\vec{n}(0)$, i.e. 
\begin{align}
\vec{n}(t)&\equiv \Upsilon(t)\left[\vec{n}(0)\right] \notag\\
    & =T(t)~\vec{n}(0)+\vec{m}(t)   \label{eq:transfer}
\end{align}
where $T(t)$ is a real matrix and $\vec{m}(t)$ is a real vector \cite{Byrd11}.
$\vec{m}(t)$ is zero for all time only when $\Upsilon(t)$ is unital, 
i.e., it maps $\vec{0}$ to $\vec 0$ (in terms of $\Lambda$, the unital property 
means that $\Lambda$ maps identity matrix to identity matrix).
$T(0) = I$ and $\vec{m}(0) = \vec{0}$.

Coherent dynamics is described by unitary transformations on the density
matrix (single Kraus operator). The dynamical map $\Upsilon(t)$ is then linear
which translates to orthogonal
transformations $T$ acting on the polarization vector $\vec{n}$. 
Decoherent dynamics (multiple Kraus operators) is characterized by the
nonorthogonality of the transfer matrix $T$. 
Markovian dynamics is conventionally defined by $\Upsilon$ possessing the semigroup property 
$\Upsilon(t_1+t_2) = \Upsilon(t_2)\Upsilon(t_1)$,
which translates to 
\begin{align}
T(t_1+t_2) =& T(t_2)T(t_1) \label{eq:T}\\
\vec{m}(t_1+t_2) =& T(t_2)\vec{m}(t_1) + \vec{m}(t_2) . \label{eq:m}
\end{align}
We shall adopt a slightly different definition of Markovianity for the present 
paper.  An evolution will be said to be ``distance Markovian" if 
$|\vec{n}(t)-\vec{n}_\infty|$ is a monotonically decreasing function.  
Note ``distance Markovian" is a weaker 
condition than Markovian, though the two are usually equivalent. 
Given the semigroup property  Eq. \ref{eq:T} and Eq. \ref{eq:m}, we have
\begin{align*}
\left|\vec{n}(t) - \vec{n}_\infty\right| &
        = \left|\left[T(t)\vec{n}(0)+\vec{m}(t)\right] 
                    - \left[T(t)\vec{n}_\infty + \vec{m}(t) \right]\right| \\
       &\le \|T(t)\| \left|\vec{n}(0)-\vec{n}_\infty \right| \\
       &\le \left|\vec{n}(0)-\vec{n}_\infty \right| 
\end{align*}
since all eigenvalues of $T(t)$ have their norms in the range $0$ to $1$, i.e.,
$T(t)$ cannot increase the purity of the quantum state.

Any model of an open quantum system defines a set of possible dynamical
evolutions. This is done by specifying the equations of motion, which give $T$, 
$\vec{m}$ and the initial conditions, which give $\vec{n}(0)$.  We
define the dynamical subspace $D$ of a model as the set of all trajectories
{allowed by the set of initial conditions and the equations of
motions}.
 Eq. \ref{eq:transfer} shows that, as long as the set of all initial
conditions is a linear space (the usual case), then $D$ is a linear space
intersected with $M$:  we first choose a basis that spans the set of all
possible $\vec{n}\left( 0\right)$, then evolve this basis according to Eq. \ref%
{eq:transfer}, giving a linear subspace in the space of all $\vec{n}$.  
A precise and general definition of $D$ is given in Appendix \ref{sec:D}.
The set of \textit{admissible} $\vec{n}$ is then given by intersecting this
linear subspace with $M$.  $D$ is a manifold of any dimension from $1$ to $15$
in the two-qubit case. 
    We note $\dim D$ could be smaller than $\dim M$. This happens when both
    $\rho(0)$ and $E_a(t)$ are expandable by identity and a true subalgebra 
    $\mathbf a$ of $su(N)$.
    $\dim D$ is then equal to the number of independent elements in $\mathbf a$.
    For example, if $\rho(0)$ is a two-qubit ``X-state" and the dynamics can be
    described by the action of Kraus operators in the X-form, the dynamical 
    subspace $D$ will be $7$-dimensional \cite{Yu07QIC}.

\textit{It is the nature of the intersection of} $D$ \textit{with} $S$ 
\textit{and the position of} $\vec{n}_{\infty}$ \textit{relative to }$S$ 
\textit{that determines the categories of entanglement evolution of a model. }

The aim of the remainder of this paper is to show how to determine the
topological structure of $D\cap S$ and the position $\vec{n}_{\infty }$ for
various illustrative models of increasing complexity, and then to deduce the
possible entanglement evolutions from this information.
We note that in general $D$ can be determined without fully solving the dynamics.
Thus it is possible to gain qualitative information of the entanglement 
evolution of the model with simple checks. 

\subsection{Model $D_{3}$}

\label{sec:su2}

Our first model consists of two qubits (A and B) with a Heisenberg interaction
and classical dephasing noise on one of the qubits.  The Hamiltonian is
\begin{equation}
H(t) = -\frac{1}{2}\left[{J}
\vec{\sigma}^A\cdot\vec{\sigma}^B+s(t){g}{\sigma}^B_z+B_0\sigma^A_z\right], 
\label{eq:H}
\end{equation}
where $s\left( t\right) $ is a random function.  This is a classical noise
model.  To compute $\vec{n}\left( t\right) $ we need to average over a
probability functional for $s\left( t\right)$, which we will specify more
precisely below. Note that the manifold spanned by $\{\ket{00},\ket{11}\}$ 
is decoupled from the manifold spanned by $\{\ket{01},\ket{10}\}$ under the 
influence of this Hamiltonian.
Thus if the initial density matrix lies in one of the two subspaces, 
the four-level problem 
decouples into two two-level problems and we can use the Bloch ball 
representation to visualize the state space and entanglement evolution.

\begin{figure}[tbp]
\centering\includegraphics*[width=.8\linewidth]{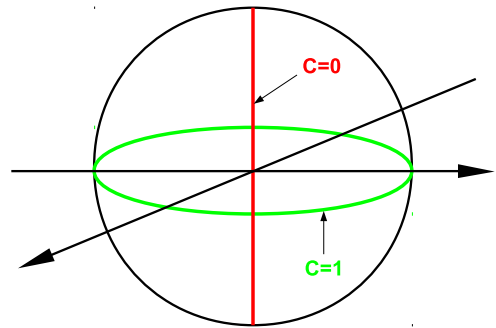}
\caption{(Color online) Effective Bloch sphere representation for the
dynamical subspace $D_3$ . The north and south pole are either $%
\{\left|10\right>,~\left|01\right>\}$ or $\{\left|00\right>,
\left|11\right>\}$. States resting on the red line connecting north and
south pole have zero concurrence. States on the green equator are equivalent
to Bell states thus maximally entangled.}
\label{fig:bloch_2}
\end{figure}

Take the initial state to be in span$\{\ket{00},\ket{11}\}$ for example.
The dynamical subspace $D_{3}$ is a $3$-dimensional ball, as
shown in Fig. \ref{fig:bloch_2}.  This makes it relatively easy to
visualize the state and entanglement evolution.  However, note that
the center of $D_{3}$ is \textit{not} the state $\vec{n} = \vec{0}$.  
In fact, all the points on the $z$-axis belong to $B_S$ 
because every neighborhood of any of these points contains points
for which $C>0$.

The square roots of the non-zero eigenvalues of $\rho \tilde{\rho}$ are $%
\lambda _{1,2}=(\sqrt{1-r^{2}\cos ^{2}\theta }\pm r\sin \theta )/{2}$, where 
$r$ and $\theta $ are the spherical coordinates of the ball, and $\tilde{\rho%
}$ is the spin-flipped density matrix, as in Eq. \ref{eq:flip}. The
concurrence is given by 
\begin{equation}
C=r\sin \theta .
\end{equation}
The maximally entangled states are on the equator and the separable states
are on the $z$-axis, as seen in Fig. \ref{fig:bloch_2}. The concurrence has
azimuthal symmetry and is linear in the radial distance from the $z$-axis.
The separable states in $D_{3}$ form the $1$-dimensional line $D_{3}\cap S$
that connects the north and south poles of $D_{3}$. 

The key point is that $D_{3}\cap S$ has a lower dimension than $D_{3}$
itself.  Now consider the possible trajectories  
with $\vec{n}_{\infty }$ on the $z$-axis.  
No function $\vec{n}(t)$ with continuous first derivative can have a finite
time interval with $C(t)=0$.  The trajectories either hit the $z$-axis at 
discrete time instants which puts them in category $\mathcal{B}$,
or approach the $z$-axis asymptotically which puts them in category 
$ \mathcal{A}$. 

Let us specify $s\left( t\right) $ in more detail to demonstrate how those
two qualitatively different behaviors are related to Markovianity.  Qubit
A sees a static field $B_{0}$ while qubit B sees a fluctuating
field $s(t)g$.  All fields are in the $z$-direction. We will take the noise
to be random telegraph noise (RTN): $s(t)$ assumes value $\pm 1$ and
switches between these two values at an average rate $\gamma $.  RTN is
widely observed in solid state systems \cite
{Kogan08,Bialczak07,Bergli09,Cheng08PRA,*Zhou10PRA}. 

For this dephasing noise model, the above-mentioned decoupling into two
2-dimensional subspaces occurs. In the $2\times 2$ block labelled by $%
\{\left\vert 00\right\rangle ,\left\vert 11\right\rangle \}$ we find 
\begin{equation*}
H(t)=-\frac{\sigma _{z}}{2}\left[ s(t)g+B_{0}\right] ,
\end{equation*}%
wherer $\sigma _{z}$ is the Pauli matrix in the subspace.

This Hamiltonian can be solved \textit{exactly} using a quasi-Hamiltonian method \cite%
{Joynt_2009}. The time-dependent decoherence problem can be mapped exactly
to a time-independent problem where the two-value fluctuating field is
described by a spin half particle. The quasi-Hamiltonian is given by 
\begin{equation*}
H_{q}=-i\gamma +i\gamma \tau _{1}+L_{z}(B_{0}+\tau _{3}g),
\end{equation*}%
where $\tau _{i}$ are the Pauli matrices of the noise ``particle". $L_{z}$ is
the $SO(3)$ generator in the $\{\left\vert 00\right\rangle ,\left\vert
11\right\rangle \}$ space.

The transfer matrix is given by 
\begin{align}
T(t) = 
\begin{bmatrix}
\zeta_{T}(t)\cos B_0t & \zeta_{T}(t)\sin B_0t & 0 \\ 
-\zeta_{T}(t)\sin B_0t & \zeta_{T}(t)\cos B_0t & 0 \\ 
0 & 0 & 1%
\end{bmatrix}%
\end{align}
where 
\begin{align}
\zeta_{T}(t) =& e^{-\gamma t} \left[\cos\left(\Omega t\right) +\frac{\gamma}{%
\Omega}\sin\left(\Omega t\right)\right]  \label{eq:zeta} \\
\Omega = &\sqrt{g^2-\gamma^2}
\end{align}
is the dephasing function due to RTN and it describes the phase coherence in
the x-y plane \cite{Zhou10QIP}. 
    $\vec{m}(t)=0$ since the dynamics is unital.
Note $\zeta_T(t)$ has qualitatively
different behaviors in the $g\le\gamma$ and $g>\gamma$ regions, 
as the trigonometric functions become hyperbolic functions \cite{Joynt_2009}.

Taking $\left|\Phi+\right>=(\left|00\right>+\left|11\right>)/\sqrt{2}$ as
initial state, the effective Bloch vector is 
\begin{align}
\vec{n}(t) = 
\begin{bmatrix}
\zeta_T(t) \cos B_0 t \\ 
- \zeta_T(t) \sin B_0 t \\ 
0%
\end{bmatrix}%
.
\end{align}
The state trajectory is fully in the equatorial plane, as seen in Fig. \ref%
{fig:xy_plane}. The dephasing function $\zeta_T(t)$ modulates the radial
variation and the static field $B_0$ provides precession. 
\begin{align}
\left|\vec{n}(t)-\vec{n}_\infty\right| = |\zeta_T(t)|,
\end{align}
and the dynamics is distance Markovian if $g\le\gamma$ ($\zeta_T(t)$ being monotonic).
In this parameter region, $\zeta_T(t)$ can be approximated by 
$\exp(-g^2/2\gamma)$ and the dynamics is approximately Markovian as well.
Thus we do not need to distinguish Markovian and distance Markovian in this 
model.  In the Markovian
case, the monotonicity of $\zeta_T$ gives rise to spiral while in the
non-Markovian case the state trajectory periodically spirals outwards with
frequency $\Omega$. In both cases, the limiting state is the origin of the ball.

\begin{figure*}[t]
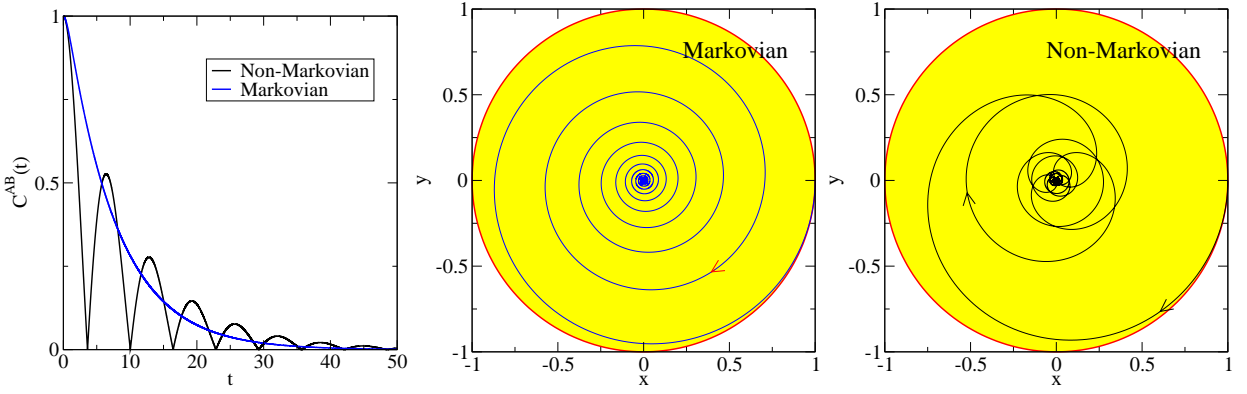

\centering
\includegraphics*[width=.3\linewidth]{zeta.eps} \includegraphics*%
[width=.3\linewidth]{M_xy.eps} \includegraphics*[width=.3 \linewidth]{NM_xy.eps}
\caption{ (Color online) (a) Entanglement evolution and (b), (c) state
trajectories in the dynamical subspace $D_3$ with RTN for Markovian and
Non-Markovian evolutions. Concurrence is equal to the absolute value of $%
\protect\zeta_T(t)$. State trajectory in the equatorial plane of the
effective Bloch representation where the only separable state is the origin $%
\vec{0}$. The initial state is the Bell state $\left|\Phi+\right>$. The
concurrence is given by the radial distance. In both cases, $g=0.5$, $B_0=1$%
. $\protect\gamma=1$ for the Markovian case and $\protect\gamma=0.1$ for the
Non-Markovian case. }
\label{fig:xy_plane}
\end{figure*}

The concurrence evolution is given by 
\begin{equation*}
C(t)=|\zeta _{T}(t)|.
\end{equation*}%
Thus Markovian noise gives rise to entanglement evolutions in category $%
\mathcal{A}$ while non-Markovian noise gives rise to that in category $%
\mathcal{B}$. Entanglement evolutions in the other two categories can not
occur due to fact that $\dim \left( D_{3}\cap S\right) <\dim \left(
D_{3}\right)$.

\subsection{Model $D_{8}$ }

\label{sec:su3}

In the previous section we saw that dynamical subspaces spanned by two
computational basis states does not possess the property $\dim D\cap S=\dim D$.
A natural question is whether simply increasing the number of
basis states helps.  This can be done by choosing a Hamiltonian that
connects only the triplet states in the original Hilbert space.  One
example would be 
\begin{equation*}
H={J(t)}~\vec{\sigma}^{A}\cdot \vec{\sigma}^{B},
\end{equation*}
as in Ref. \cite{Testolin09}, where the Heisenberg coupling $J(t)$ has time dependence 
and is modeled as a classical random process.

Note this Hamiltonian conserves the total angular momentum of the two qubits.
As a result, the triplet space 
spanned by $\{\left\vert 00\right\rangle ,\left\vert
11\right\rangle ,\left( \left\vert 10\right\rangle +\left\vert
01\right\rangle \right) /\sqrt{2}\}$ is decoupled from the singlet space. 
Thus the dynamical subspace $D$ has its basis elements in $su(3)$ if 
we choose the initial state to be in the triplet subspace.

Using Gell-Mann matrices as the elements of $su(3)$ algebra, the state space
is a subset of a ball in $\mathbb{R}^{8}$ \cite{Kimura03}. 
\begin{equation*}
\rho =\frac{1}{2}I+\frac{1}{2}\sum_{i=1}^{8}m_{i}\mu _{i}
\end{equation*}%
and $\mu _{i}$ satisfies 
\begin{equation*}
\mu _{i}=\mu _{i}^{\dagger },\quad \text{Tr~}\mu _{i}=0,\quad \text{Tr~}\mu
_{i}\mu _{j}=2\delta _{ij}.
\end{equation*}%
The $m_{i}$ are linearly related to the previously defined $n_{i}$.

The square roots of the eivenvalues of $\rho\tilde\rho$ are 
\begin{widetext}
\begin{align}
\lambda_{1,2}=\left|\frac{1}{2}\sqrt{m_{6}^2+m_{7}^2}\pm\frac{\sqrt{2}}{6}\sqrt{2-\sqrt{3}\left(\sqrt{3}m_{3}+m_{8}\right)+3m_{8}\left(\sqrt{3}m_{3}-m_{8}\right)} \right|, \ \ \lambda_{3,4}=0
\end{align}
\end{widetext}

Thus the set $D_{8}\cap S$ of separable states is composed of two geometric
objects: $m_{6}=m_{7}=0$ which is in $\mathbb{R}^{6}$ and 
\begin{equation*}
2-\sqrt{3}\left( \sqrt{3}m_{3}+m_{8}\right) +3m_{8}\left( \sqrt{3}%
m_{3}-m_{8}\right) =0
\end{equation*}%
which is in $\mathbb{R}^{7}$. In addition to a concurrence-zero hyperline, we
have a $C=0$ hyperplane in $D_{8}$.  Hence$~\dim \left( D_{8}\cap S\right)
<\dim \left( D_{8}\right) $ and this model only displays entanglement
evolutions in categories $\mathcal{A}$ and $\mathcal{B}$.

Introducing extra dimensions thus helps to form non-zero volume of $D\cap S$ in $D$
but a Hilbert space spanned by three of the four computational
basis states is still not enough. $D_{3}$ and $D_{8}$ both avoid the region
near the fully mixed state $\vec{0}$, where most separable states reside 
\cite{Gurvits03,*Gurvits05}. This region is included in the dynamical
subspaces in the next two sections.

Note that for $D_3$ and $D_8$, the symmetry of the Hamiltonian and the specification
of the initial conditions allow us to fully describe the dynamical subspace without 
explicitly solving the dynamics. 
This feature can be seen in the more complicated models in the 
following sections as well: entanglement evolution categories, as a qualitative 
property of the system dynamics, can be determined from symmetry considerations
of the model (dynamics plus initial condition), position of $\vec{n}_\infty$ in 
$D\cap S$ and the memory effect of the environment.

\subsection{Model $YE$}

\label{sec:YE}

Yu and Eberly considered a disentanglement process due to spontaneous
emission for two two-level atoms in two cavities.  In this case the
decoherence clearly acts independently on the two qubits.  Nevertheless they
found that ESD occurs for specific choices of initially entangled states 
\cite{Yu04PRL}. 

The decoherence process is formulated using the Kraus operators
\begin{align}
\rho(t) = \sum_{\mu=1}^4K_\mu(t)\rho(0)K_\mu^\dagger(t), \label{eq:ye}
\end{align}
where $K_\mu(t)$ satisfy $\sum_\mu K^\dagger_\mu K_\mu=I$
for all $t$ \cite{Kraus83,Nielsen00Chuang}. For the atom-in-cavity model,
they are explicitly given by 
\begin{align}
K_1 =& F_1\otimes F_1,\quad K_2 = F_1 \otimes F_2 \\
K_3 =& F_2\otimes F_1, \quad K_4 = F_2 \otimes F_2,
\end{align}
where 
\begin{align}
F_1=%
\begin{bmatrix}
\gamma & 0 \\ 
0 & 1%
\end{bmatrix}%
, F_2 = 
\begin{bmatrix}
0 & 0 \\ 
\omega & 0%
\end{bmatrix}%
,
\end{align}
and $\gamma=\exp(-\Gamma t/2)$ and $\omega=\sqrt{1-\exp(-\Gamma t)}$.

It is possible to choose initial states such that the density matrices have
the following form for all $t$ 
\begin{align}
\rho(t) = \frac{1}{3} 
\begin{bmatrix}
a(t) & 0 & 0 & 0 \\ 
0 & b(t) & z(t) & 0 \\ 
0 & z(t) & c(t) & 0 \\ 
0 & 0 & 0 & d(t)%
\end{bmatrix}%
\end{align}
with 
\begin{align}
a(t) &= \kappa^2 a_0, \\
b(t) &= c(t) = \kappa + \kappa(1-\kappa)a_0, \\
d(t) &= 1-a_0+2(1-\kappa)+(1-\kappa)^2 a_0, \\
z(t) &= \kappa.
\end{align}
where $\kappa = \exp(-\Gamma t)$.
Here the parameter $a_0$ determines the initial condition.

The two-qubit entanglement is 
\begin{align}
C(t) = \frac{2}{3} \max\left\{0,\kappa f(t)\right\},
\end{align}
where $f(t) = 1-\sqrt{a_0[1-a_0+2(1-\kappa)+(1-\kappa)^2a_0]}$.

In the polarization vector representation, the dynamics
defined by Eq. \ref{eq:ye} can be given explicitly 
by the transfer matrix $T(t)$ and the translation vector $\vec{m}(t)$
\begin{align}
T(t) = \begin{pmatrix}
    \kappa & 0 & 0 & 0 & 0 \\
    0 & \kappa & 0 & 0 & 0 \\
    0 & 0 & \kappa & 0 & 0 \\
    0 & 0 & 0 & \kappa  & 0 \\
    \kappa^2-\kappa & 0 & 0 & \kappa^2-\kappa  & \kappa^2 \end{pmatrix},
\end{align}
and 
\begin{align}
\vec{m}(t) = [\kappa-1;0;0;\kappa-1;(\kappa-1)^2].
\end{align}
Here the coordinates are $\{n_{IZ},n_{XX},n_{YY},n_{ZI},n_{ZZ}\}$.
We note that Eqs. \ref{eq:T} and \ref{eq:m} are satisfied and this spontaneous 
emission model is Markovian and distance Markovian.

The non-zero components are 
\begin{align}
n_{IZ}(t)=& n_{ZI}(t)=-1+\frac{2}{3}(1+a_{0})\kappa \\
n_{XX}(t)=& n_{YY}(t)=\frac{2}{3}\kappa \\
n_{ZZ}(t)=& 1-\frac{4}{3}(1+a_{0})\kappa +\frac{4}{3}a_{0}\kappa ^{2}.
\end{align}%
and the limiting state is 
\begin{equation}
\vec{n}_{\infty }=(-1,0,1),
\end{equation}
where the coordinates are $(n_{IZ},n_{XX},n_{ZZ})$.

This shows that the dynamical subspace $D_{Y}$ is a 3-dimensional section of 
$M$ where the non-zero components are $n_{IZ}$,$n_{ZI},n_{XX},n_{YY},$ and $%
n_{ZZ}$ but also $n_{IZ}=n_{ZI}$ and $n_{XX}=n_{YY}$,  such that it can
also be visualized in three dimensions. Interestingly, the limiting state 
$\vec{n}_{\infty }$ due to spontaneous emission is on the boundary of set $S$
of separable states, and the purity of the state increases with time.

Although we have so far fully solved the system dynamics, for the purpose of describing
$D_Y$, it is enough to know that $\{IZ, ZI, XX, YY,ZZ\}$ are the basis of $D_Y$ and that
the Kraus operators preserve the equalities $n_{IZ}=n_{ZI}$, $n_{XX}=n_{YY}$ from 
the initial conditions.
$D_Y$ is then determined from the positivity condition of the density matrix.

The positivity condition for the density matrix is given by 
\begin{align*}
a_2\ge0&\Rightarrow 2n_{IZ}^2+2n_{XX}^2+n_{ZZ}^2\le3 \\
a_3\ge0&\Rightarrow 2n_{IZ}^2+2n_{XX}^2+n_{ZZ}^2\le
1+2n_{ZZ}\left(n_{IZ}^2-n_{XX}^2\right) \\
a_4\ge0&\Rightarrow AB\ge0
\end{align*}
where $A = (1+n_{ZZ})^2-4n_{IZ}^2$ and $B=(1-n_{ZZ})^2-4n_{XX}^2$.
At $t=0$, $\vec{n}_0= (2a_0-1,2,-1)/3$. The positivity constraint $a_3\ge0$
gives rise to the range of the possible initial conditions, parametrized by  
$a_0 \in[0,1]$.

The square roots of the eigenvalues of $\rho \tilde{\rho}$ are 
\begin{align}
\lambda _{a}=& \lambda _{b}=\frac{1}{4}\sqrt{A} \\
\lambda _{c}=& \frac{1}{4}\left\vert 1-n_{ZZ}+2n_{XX}\right\vert \\
\lambda _{d}=& \frac{1}{4}\left\vert 1-n_{ZZ}-2n_{XX}\right\vert
\end{align}%
The ordering of the $\lambda $'s can change during the course of an
evolution.  When $\lambda _{a}$ is the largest one finds $C=0,$ which is
helpful in dtermining $B_S$.

\begin{figure}[tb]
\centering\includegraphics*[width=.8\linewidth]{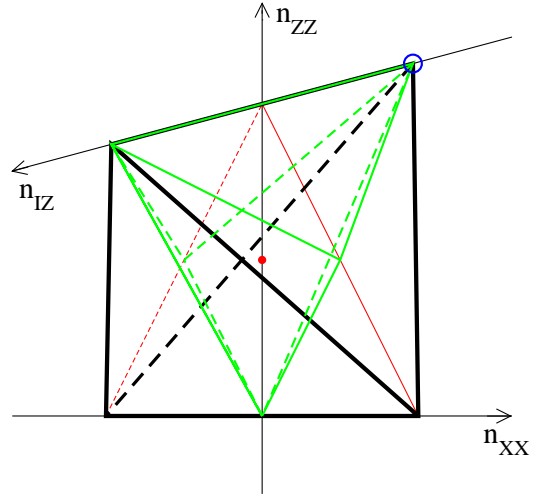}
\caption{(Color online) The dynamical subspace $D_Y$ is a tetrahedron
outlined by black lines. The set of separable states $D_Y\cap S$ form a
hexahedron outlined by green lines. The origin (fully mixed state $\vec{n}=%
\vec{0}$) is denoted by a red filled dot. The limiting state $\vec{n}%
_\infty$ in Ref. \protect\cite{Yu04PRL} is denoted by a blue circle. }
\label{fig:sig5}
\end{figure}

$D_{Y}$ is a tetrahedron with vertices at $(0,-1,-1)$, $(0,1,-1)$, $(-1,0,1)$%
, and $(1,0,1)$, as seen in Fig. \ref{fig:sig5}. $D_{Y}\cap S$ is a
hexahedron that shares some external areas with $D_{Y}$.  The $2$-dimensional
section of $D_{Y}$ with $n_{IZ}=0$ is shown in Fig. \ref%
{fig:contour}. On the other hand, if the section is done with $n_{XX}=0$, we
get a upside down triangle made of separable states, as shown in Fig. \ref%
{fig:sec_n3}.

\begin{figure}[tb]
\centering\includegraphics[width=.8\linewidth]{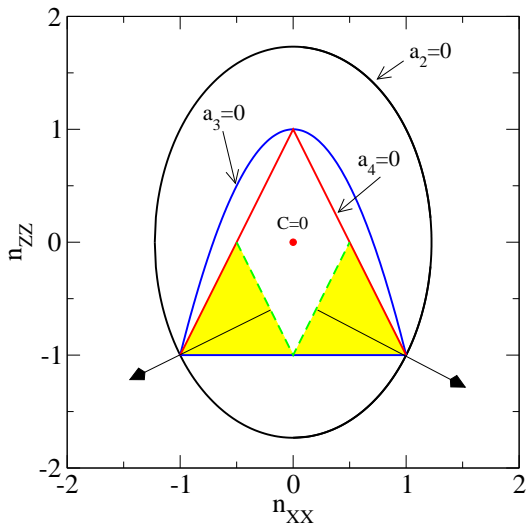}
\caption{(Color online) Cross section of the dynamical subspace $D_Y$ with
$n_{IZ}=0$ and $D_Z$ with $n_{XY}=0$. 
$a_2=0$ is the ellipse. $a_3=0$ gives the parabola and the
bottom of the isosceles triangle. $a_4=0$ sets the two sides of the
triangle. The entangled region is shaded where the filled arrows denote
increasing direction of the concurrence. The green dashed line is the
boundary of entangled and separable states. The fully mixed state is
denoted by a red dot. }
\label{fig:contour}
\end{figure}

\begin{figure}[tb]
\centering\includegraphics*[width=.8\linewidth]{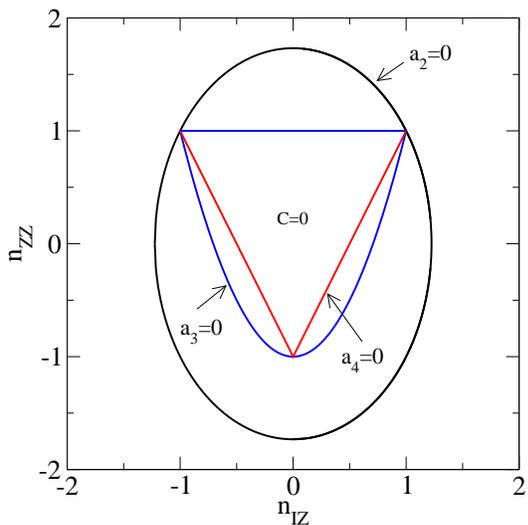}
\caption{(Color online) Cross section of the dynamical subspace $D_Y$ with $%
n_{XX}=0$. $a_2=0$ is the ellipse. $a_3=0$ gives the parabola and the
top of the isosceles triangle. $a_4=0$ sets the two sides of the triangle.
The whole section is filled with separable states. }
\label{fig:sec_n3}
\end{figure}

\begin{figure}[tbp]
\centering\includegraphics*[width=0.8\linewidth]{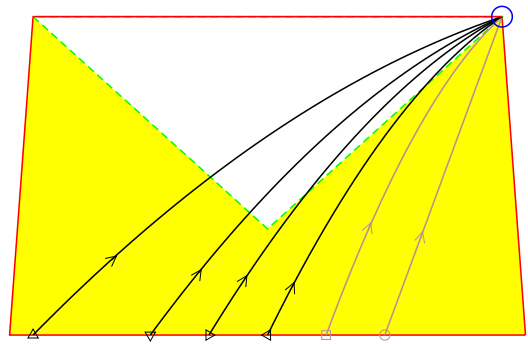}
\caption{(Color online) Skematic drawing of the different types of
entanglement evolutions in the dynamical subspace $D_Y$. The limiting state 
$\vec{n}_\infty$ is denoted by a blue circle on the top right corner.
Entanglement evolutions in category $\mathcal{A}$ are brown-colored while
those in category $\mathcal{E}$ are black-colored. }
\label{fig:evo}
\end{figure}

Yu and Eberly showed that a sudden transition of the entanglement evolution
from category $\mathcal{A}$ to category $\mathcal{E}$ is possible as one
tunes the physical parameter $a_{0}$. This phenomenon can be
easily understood in our formalism, as seen in Fig. \ref{fig:evo}. The
curvature of the state trajectories vary as the initial state changes. Thus
there is a continuous range of initial states parametrized by $a_{0}$ whose
trajectories enter $S\cap D_Y$ within finite amount of time and also a continuous
range of initial states whose trajectories never enter Int$(S)$. Note Fig. 
\ref{fig:evo} is a schematic drawing since the true state trajectories are
truly three dimensional.

To be more quantitative, the transition between the category $\mathcal{A}$ and 
category $\mathcal{E}$ behaviors in the YE model could be determined by 
examining the angle $\theta_\infty$ between $B_S$ and the tangent vector of 
the state trajectory in the long time limit.
We denote the limiting tangent vector by $\vec{n}_T(\infty)$ and it is given by
\begin{align}
\vec{n}_T(\infty) = (1+a_0,1,4a_0-2).
\end{align}

The relevant $B_S$ in the YE model is a plane passing defined by the following 
three points: $\vec{n}_\infty$, $(0,0,-1)$ and  $(0,1/2,0)$. 
It is parametrized by
\begin{align}
\hat{m} \cdot ( \vec{n} - \vec{n}_\infty) =0,
\end{align}
where $\hat{m}=(2,-2,1)/3$ is the unit normal of the plane pointing into the 
separable region $S\cap D_Y$.
Note $\vec{m} \cdot \vec{n}_T(\infty)=(6a_0-2)/3$ is proportional to 
$\cos \theta_\infty$ and its sign tells us whether the state trajectory 
approaches $\vec{n}_\infty$ from the separable region or the entangled region.
Since $0\le a_0\le 1$, $\vec{m} \cdot \vec{n}_T(\infty)$ falls in the range
$[-2/3,4/3]$ and both the ESD and HL behaviors are possible.

The condition
\begin{align}
\vec{m} \cdot \vec{n}_T(\infty)=0,
\end{align}
i.e., $a_0=1/3$, gives rise to the critical trajectory which approaches 
$\vec{n}_\infty$ along $B_S$.
When $\vec{m} \cdot \vec{n}_T(\infty)>0$, i.e., $a_0\in[1/3,1]$, the state 
trajectory approaches $\vec{n}_\infty$ from the entangled region and we get
entanglement evolutions in category $\mathcal A$. These trajectories are 
represented by the brown curves in Fig. \ref{fig:evo}.
On the other hand, when $\vec{m} \cdot \vec{n}_T(\infty)\le0$, i.e., 
$a_0\in[0,1/3)$, the state trajectory approaches $\vec{n}_\infty$ from the 
separable region and we get entanglement evolutions in category $\mathcal E$.
These trajectories are represented by the black curves in Fig. \ref{fig:evo}.

The key point about the model $YE$ is that the limiting state $\vec{n}%
_{\infty }\in $ $B_S$: it is on the boundary of the entangled
and separable regions.  That is why entanglement evolutions in both
category $\mathcal{A}$ and category $\mathcal{E}$ are possible.

\subsection{Model $ZJ$}

\label{sec:ZJ} 

Here we present a physically motivated dynamical subspace
where the dynamics satisfies the following conditions: 1) the two qubits are
not interacting; 2) the noises on the qubits are not correlated; 3) the
effect of dephasing and relaxation can be separated. This model shows all four
categories of entanglement evolution. 

For this model the two-qubit dynamics can be decomposed into single-qubit
dynamics \cite{Zhou10QIP}. The extended two-qubit transfer matrix is $%
\underline{T}(t)=\underline{R}^{A}\otimes \underline{R}^{B}$ where 
\begin{equation*}
\underline{R}(t)=%
\begin{bmatrix}
1 & 0 & 0 & 0 \\ 
0 & \zeta \left( t\right) \cos B_{0}t & \zeta \left( t\right) \sin B_{0}t & 0
\\ 
0 & -\zeta \left( t\right) \sin B_{0}t & \zeta \left( t\right) \cos B_{0}t & 
0 \\ 
0 & 0 & 0 & e^{-\Gamma _{1}t}%
\end{bmatrix}%
\end{equation*}%
is the extended transfer matrix of individual qubits. The top left entry $1$
describes the dynamics of $I$ and is there only for notational convenience.
Here $\zeta (t)$ describes dephasing process, $\Gamma _{1}$ is the
longitudinal relaxation rate and $B_{0}$ is static field in the $z$
direction that causes Larmor precession. $\zeta (0)=1$ and $\zeta (\infty
)=0 $ if dephasing occurs. Note this dynamical description of decoherence is
completely general as long as one can separate dephasing and relaxation
channels.

The dynamical subspace in this model is a specially parametrized $3$-dimensional
section of the full two-qubit state space $M$. Only the components $n_{XX}$, 
$n_{XY}$, $n_{YX}$, $n_{YY}$ and $n_{ZZ}$ are non-zero and we further have
constraints $n_{XX}=n_{YY}$ and $n_{XY}=-n_{YX}$. We thus use $n_{XX,XY,ZZ}$
as independent parameters and the state space can be visualized in three
dimensions. This dynamical subspace has been previously considered in Ref. 
\cite{Zhou11QIP} and we will call it $D_{Z}$.

The fact that $\dim D<\dim M$ relies on judicious choice of the initial states $\rho_0$.
In Ref. \cite{Zhou10QIP}, more general initial states are considered such that 
$D_Z$ is expanded into a dynamical subspace with $7$ elements in the $su(4)$ algebra, 
i.e., $\{IX, XX, XY, YX, {YY}, ZI, ZZ\}$.

Note $n_{XX}^{2}+n_{XY}^{2}=R^{2}$ is conserved in $D_{Z}$. The positivity
of the density matrix requires 
\begin{align}
a_{2}\geq 0& ~\Rightarrow ~2R^{2}+n_{ZZ}^{2}\leq 3  \label{eq:a2} \\
a_{3}\geq 0& ~\Rightarrow ~n_{ZZ}\leq 1-2R^{2},~\text{and}~n_{ZZ}\geq -1 \\
a_{4}\geq 0& ~\Rightarrow ~2R+n_{ZZ}\leq 1
\end{align}%
The concurrence is given by 
\begin{equation}
C=\max \left\{ 0,R-\frac{1+n_{ZZ}}{2}\right\} .
\end{equation}

Separable states form a spindle shape on top and entangled states form a
torus-like shape on bottom with triangular cross sections. A section along $%
n_{XY}=0$ is shown in Fig. \ref{fig:contour}.

We have thus fully described the entanglement topology of $D_{Z}$ and now we
construct entanglement evolutions that induce $D_{Z}$. A model similar to
that of Eq. \ref{eq:H} that satisfies the three conditions is 
\begin{equation}
H(t)=-\frac{1}{2}\left[ s(t)\vec{g}\cdot \vec{\sigma}^{B}+B_{0}\sigma
_{z}^{A}\right].
\label{eq:zj5}
\end{equation}
Note the RTN has both dephasing ($g_{z}$) and relaxation ($g_{x},g_{y}$)
effects in this case. Situations with $n_{XY}=0$ at intermediate qubit
working point (with the presence of both dephasing and relaxation noise)
have been considered in Ref. \cite{Zhou11QIP}. Here we choose the initial
state to be the generalized Werner state \cite{Werner89} 
\begin{equation}
w_{r}^{\Phi }=r\ket{\Phi}\bra{\Phi}+ \frac{1-r}{4}I_{4},
\end{equation}
where 
\begin{align}
\ket{\Phi}=\frac{1}{2}(\ket{00} +e^{i\phi }\ket{11})
\end{align}
is a Bell state.

The state trajectory is then given by 
\begin{align}
n_{XX}(t)=& r\cos(2B_0t)\zeta(t), \\
n_{XY}(t)=& -r\sin(2B_0t)\zeta(t), \\
n_{ZZ}(t)=& r e^{-\Gamma _{1}t}.
\end{align}

Similarly, if the Werner state derived from the Bell state 
\begin{align}
\left|\Psi\right> =\frac{1}{2}\left(\left|01\right>+e^{i\phi}\left|10\right>%
\right)
\end{align}
is used as initial state, the state trajectory is 
\begin{align}
n_{XX}(t)=& r\zeta_T(t), \\
n_{ZZ}(t)=& -r e^{-\Gamma _{1}t}.
\end{align}

In both cases, the evolution of the concurrence is 
\begin{equation*}
C(t)=\max \{0,r\left[ \left\vert \zeta(t)\right\vert -\xi (t)\right] \}
\end{equation*}%
where $\xi (t)=(1-re^{-\Gamma _{1}t})/{2}$. Thus dephasing (relaxation)
drives the state horizontally (vertically) towards the origin. Note ESD
occurs whenever $\xi (t\rightarrow \infty )\neq 0$ and the corresponding
limiting state is $\vec{n}_{\infty }=\vec{0}$ \cite{Zhou10QIP}. In the
case of pure dephasing, $\Gamma _{1}=0$ and the trajectory moves in the
horizontal plane $n_{ZZ}=-r$. The limiting state is $\vec{n}_{\infty
}=(0,0,-r)$. The only case where ESD does not happen is the pure dephasing
processes on the $n_{ZZ}=-1$ plane since $\vec{n}_{\infty }\in $ $B_S$. 
 This amounts to $\xi (t)=0$ and there is no abrupt cut-off.

At the pure dephasing point, {$|\vec{n}(t)-\vec{n}_\infty|=\sqrt{2}r|\zeta(t)|$
and the dynamics is distance-Markovian if $\zeta(t)$ is monotonic.
Since the typical Markovian dynamics can be characterized by $\zeta(t)$ with 
exponential decay forms, we do not distinguish Markovian and distance-Markovian
for this model.
 Entanglement evolutions in
categories $\mathcal{E}$ and $\mathcal{O}$ are possible, depending on
whether the RTN is Markovian or non-Markovian, as seen in Fig. \ref%
{fig:pd_zj}.

\begin{figure*}[ht]
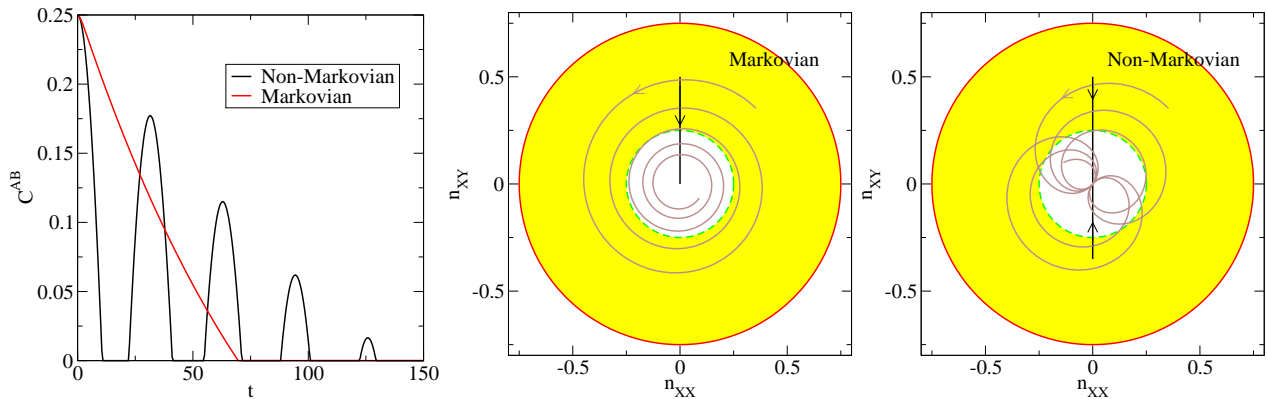

\centering
\includegraphics*[width=.32\linewidth]{pd_section_C.eps} %
\includegraphics*[width=.3\linewidth]{pd_section_M.eps} %
\includegraphics*[width=.3\linewidth]{pd_section_NM.eps}
\caption{(Color online) (a) Entanglement evolution and (b), (c) state
trajectories in $D_Z$ with pure dephasing noise. Only one qubit is subject
to RTN and the other sees a static field $B_0$. State trajectories with
initial state $w_r^{\Psi+}$ is colored brown while those with initial state $%
w_r^{\Phi+}$ are colored black. 
$\zeta(t)$ is given by Eq. \ref{eq:zeta} and $\gamma=0.5$, $g=0.1$, $r=0.5$, $
B_0=0.1$.}
\label{fig:pd_zj}
\end{figure*}

If the initial state is on the $n_{ZZ}=-1$ plane, by tuning the relative
weight of dephasing and relaxation noise, the entanglement evolution can
transit from category $\mathcal{A}$ to category $\mathcal{E}$ if the noise
is Markovian, or from category $\mathcal{B}$ to category $\mathcal{O}$ if
the noise is non-Markovian \cite{Zhou11QIP}. These transitions can be
interpreted as a transition of entanglement topology of the dynamical
subspace as the physical parameter varies. At the pure dephasing point, $%
D_{Z}$ shrinks to the $n_{ZZ}=-1$ disc and $\vec{n}_{\infty }\in $ $B_S$. 
 When this occurs we find $\mathcal{A}$ and $\mathcal{B}$
behavior.

\section{ evolution categories}
\label{sec:Ecat}

Having given examples of models that produce the most important categories
of entanglement evolution, we now provide a proof that the four categories
essentially exhaust the possibilities.  There are three cases.

Case 1. $\vec{n}_{\infty }\in \text{Int}(S)$, Image$\left( \vec{n}\right) $ and $
B_{S}$ are transversal everywhere. 
    Here $\vec{n}: [0,\infty)\longrightarrow M$ is considered as a map from the 
    time domain to the state space.

Consider a single trajectory $\vec{n}\left( t\right)$.  
  The $1$-dimensional manifold Image $\left( \vec{n}
\right) \subset M$ and the $14$-dimensional manifold $B_{S}$ satisfy $\dim
(M)=\dim \left[ \text{Image}\left( \vec{n}\right) \right] +\dim \left(
B_{S}\right)$. Hence we can apply mod $2$ intersection theory to the
intersections of these two manifolds \cite{Pollack74}.  First assume that 
{Image}$\left( \vec{n}\right) $ and $\left( B_{S}\right) $ are transversal
everywhere.  (We discuss this assumption further below.) Since $M$ is
convex, we are guaranteed that there is a straight-line path from $\vec{n}
\left( 0\right) $ to $\vec{n}_{\infty }$ and that the actual $\vec{n}\left(
t\right) $ is homotopic to this path.  The number of times that the
straight-line path crosses $B_{S}$ is one. The mod $2$ intersection theorem
then states that the number of times that $\vec{n}\left( t\right) $ crosses $
B_{S}$ is odd (formally, this is referred to as the cardinality of the
0-dimensional manifold $\vec{n}^{-1}\left( B_{S}\right) )$.  This yields two
categories as we have defined them: {card}$~\{\vec{n}^{-1}\left( B_{S}\right)
\}=1$ gives category $\mathcal E$, while {card}$~\{\vec{n}^{-1}\left( B_{S}\right)
\}=3,5,...$ gives category $\mathcal O$.

Case 2. $\vec{n}_{\infty }\in B_{S}$, Image$\left( \vec{n}\right) $ and $
B_{S}$ are transversal everywhere.

We are again assured of the existence of the straight-line path.  However,
this path can belong either to category $\mathcal A$: the path does not intersect 
$B_{S}$ and the approach to $\vec{n}_{\infty }$ is from $M\backslash S$, or to 
category $\mathcal E$: the path intersects $B_{S}$ once, and the approach to 
$\vec{n}_{\infty }$ is from $S$.  Because $S$ is convex, multiple
intersections of the straight-line path with $B_{S}$ are not allowed.  All
other paths are again homotopic to the straight-line path.  The
intersection theorem implies that if a general path approaches $\vec{n}
_{\infty }$ from $M\backslash S$, then it belongs to category $\mathcal A$ or 
$\mathcal O$ (in the variant where $C\left( t\right) >0$ for $t>t_{f}$, 
for some finite $t_{f}$), while if the path approaches $\vec{n}_{\infty }$ from $S$,
then it belongs to category $\mathcal A$ or $\mathcal O$ (in the variant where 
$C\left( t\right) =0$ for $t>t_{f}$).

Case 3. Image$\left( \vec{n}\right) $ and $B_{S}$ are not transversal
everywhere. 

There is the possibility that Image$\left( \vec{n}\right)$ and $B_{S}$
are not transversal at discrete values of $t$.  This gives rise to
quadratic or higher-order zeros of $C\left( t\right) $; the occurrence of
such points is not of great interest since they lie in a set of measure zero
that is not related to physical conditions.  Of greater interest is the
case when non-transversal points of intersection are due to restrictions
placed on the paths that stem from physical restrictions placed on a model,
e.g., a symmetry.  For a given model we must then consider $D$, its
dynamical subspace (a precise description of $D$ is given in Appendix \ref{sec:D}).
 We must replace $M$ by $D$, $S$ by $D\cap S$, and $B_{S}$ by $B_{S}\cap D$.
 The dimensions of these spaces are model-specific.  For mod $2$
intersection theory to apply, we need $\dim \left( D\right) =\dim \left[
\text{Image}\left( \vec{n}\right) \right] +\dim \left( B_{S}\cap D\right)
=1+[\dim (S\cap D)-1]=\dim (S\cap D)$.  If this holds, then the reasoning applied
to cases 1 and 2 can be repeated.
However, it is also possible to have  $\dim (S\cap D)<\dim \left( D\right) $
- we have seen an example in Fig. \ref{fig:bloch_2}.  
In this case the intersections of  Image$\left( \vec{n}\right) $ and $B_{S}\cap D$
are never transversal.  A generic path $\vec{n}\left( t\right) $ must either 
approach or ``pierce" the low-dimensional space.  This gives rise to $\mathcal A$ 
and $\mathcal B$ behaviors, respectively.

\section{results and conclusions}

\label{sec:conc}

We now summarize the results exemplified in Sec. \ref{sec:dyn} and formalized 
in Sec. \ref{sec:Ecat} by outlining a protocol to determine entanglement 
dynamics of any physical model.     

For any physical model, the possible entanglement evolution categories are 
determined by three factors:   the limiting point $\vec{n}_\infty$, the 
dynamical subspace $D$, and the presence or absence of distance-Markovianity.
Once these questions are resolved, the possible entanglement evolutions are 
given by Table \ref{tab:topo}.  

\begin{table}[tb]
\centering
\begin{tabular}{c||c|c}
\hline
& $\dim (D\cap S) = \dim D$ & $\dim (D\cap S) < \dim D$ \\ \hline\hline
$\vec{n}_\infty \in \text{Int}(S)$ & $\mathcal{E}$ or $\mathcal{O}$ & %
$\mathcal A$ or $\mathcal B$ \\ \hline
$\vec{n}_\infty \in B_S$ & $\mathcal{A}$, $\mathcal{B}$, $\mathcal{E}$ or $\mathcal{O}$ & $%
\mathcal{A}$ or $\mathcal{B}$ \\ \hline
\end{tabular}%
\caption{The relationships between entanglement evolution categories and details of 
the dynamics, such as the position of the limiting state $\vec{n}_\infty$, etc.}
\label{tab:topo}
\end{table}

$\vec{n}_\infty$ is the fixed point of the asymptotic dynamics.  Since it is 
determined by the long-time limit of the dynamics, it is usually fairly easy to
compute.  It may be a function of the parameters of the model.  The crucial 
question about $\vec{n}_\infty$ is whether it belongs to Int$(S)$,  the 
interior of $S$ (Row 1 of Table \ref{tab:topo}) or $B_S$, the boundary of $S$ 
(Row 2 of Table \ref{tab:topo}).  $D$ is the collection of all trajectories of the 
model.  It depends on all parameters of the model, and on their range of 
variation.   A prescription for computing $D$ has been given in Sec. 
\ref{sec:dyn} A.  The crucial questions about $D$ are its dimensionality $\dim(D)$ 
and the dimensionality of $\dim(D\cap S)$.  If $\dim(D) = \dim(D\cap S)$ then 
Column 1 of Table \ref{tab:topo} applies.  If $\dim(D) <\dim(D\cap S)$ then Column 
2 of Table \ref{tab:topo} applies.  Finally, non-distance-Markovian evolution gives 
$\mathcal B$ and $\mathcal O$, while distance-Markovian evolution typically gives 
$\mathcal E$ and $\mathcal A$.

Once the possible behaviors are known, one would like also to be able to 
compute the critical points or surfaces where transitions from one to another 
occur.  

The first type of transition occurs when $\vec{n}_\infty$ does not move between
$B_S$ and Int$(S)$, i.e., we stay in one row of the table.  One cannot move 
between columns within a given model since $D$, by definition, represents all 
possible states of a model.   Thus we stay in one box of the table. 
$\mathcal E \longleftrightarrow \mathcal O$ and $\mathcal A \longleftrightarrow
\mathcal B$ are the possible transitions and the critical points represent a 
change from distance-Markovianity to non-distance-Markovianity. Such transitions
may be subtle and difficult to calculate, and an explicit equation for the 
critical surface cannot be given.  A quite simple example where the calculation
can be done analytically has been given in Sec.  \ref{sec:dyn} B.  

The second type of transition happens when $\vec{n}_\infty$ moves between $B_S$
and Int$(S)$ and $\dim(D) = \dim(D\cap S)$ (Column 1).  There will be a specific
point at which this occurs.  This point is determined by examining the 
asymptotic dynamics of the model as a function of the parameters.  If the 
parameters are collectively given by a set $\{g_i\}$, then the equation for the 
critical surface is $\vec{n}_\infty(g_i)\in B_S$. The transitions are $\mathcal 
E \longleftrightarrow \mathcal A$ and $\mathcal O\longleftrightarrow \mathcal B$ for
distance-Markovian and non-distance-Markovian evolution. 
An example has been given in Sec. \ref{sec:dyn} E and Ref. \cite{Zhou11QIP}.

The third type of transition occurs in Column 1, Row 2 of the table.  In this 
case $\vec{n}_\infty$ does not change, but the direction of approach to 
$\vec{n}_\infty$ changes as a function of the parameters.  In this case the 
equation of the critical surface is $\vec{n}_T(\infty)\in \text{tang}(B_S)$, where 
$\vec{n}_T(\infty)$ is the limiting tangent vector of the state trajectory and 
tang$(B_S)$ is the tangent space of the boundary of $S$.  The transitions are 
$\mathcal E \longleftrightarrow \mathcal A$ and 
$\mathcal O \longleftrightarrow\mathcal B$ for distance-Markovian and 
non-distance-Markovian evolution.  An example has been given in Sec. 
\ref{sec:dyn} D. 

``Tricritical" points can also occur when more than one critical condition is 
satisfied.

\appendix
\section{two dimensional sections}
\label{sec:section} 

With our conventions for the two-qubit $su(4)$ algebra, 
the structure constants can be calculated from 
\begin{align}
f_{ijk}  =& \frac{1}{2i\times4}\text{Tr~}\left[\mu_i,\mu_j\right] \mu_k\\
d_{ijk}  =& \frac{1}{2\times4}\text{Tr~}\left[\mu_i,\mu_j\right]_+ \mu_k,
\end{align}
Here we denote $[,]_+$ as the anti-commutation operation.
Most of them are zero and the non-zero structure constants are  
\begin{align*}
1 =&d_{1,4,5} = d_{1,8,9} = d_{1,12,13} = d_{2,4,6} = d_{2,8,10} \\
 =&  d_{2,12,14} =d_{3,4,7} = d_{3,8,11} = d_{3,12,15} =-d_{5,10,15} \\
  =& d_{5,11,14}= d_{6,9,15}  = -d_{6,11,13} = -d_{7,9,14} = d_{7,10,13} \\
 =& f_{1,2,3} = f_{1,6,7} = f_{1,10,11}= f_{1,14,15} = -f_{2,5,7} \\
 =& -f_{2,9,11} = -f_{2,13,15} = f_{3,5,6}= f_{3,9,10}=f_{3,13,14} \\
 =& f_{4,8,12} = f_{4,9,13} = f_{4,10,14} = f_{4,11,15} = f_{5,8,13}\\
 =& f_{5,9,12} = f_{6,8,14} = f_{6,10,12} = f_{7,8,15} = f_{7,11,12}.
\end{align*}

Suppose the two non-zero components of a two-dimensional section assume 
numerical values $x$ and $y$.  It can be shown that 
\begin{align}
\text{Tr~} \rho^2 =& \frac{1}{4}(1+R^2) \\
\text{Tr~} \rho^3 =& \frac{1}{16}(1+3R^2) \\
\text{Tr~}\rho^4 =& \frac{1}{64}[1+6R^2+R^4 + n_in_jn_\ell n_m d_{ijk}d_{k\ell m}],
\end{align}
where $R^2 = x^2+y^2$.
Note the generators $\mu_i$ as defined in Eq. \ref{eq:pauli} are elements of the 
Pauli group thus they either commute or anti-commute. 
This is the origin of the two geometrically different $2$-dimensional sections in $M$.

If the two $\mu_i$'s anti-commute, the $d_{ijk}$ related term vanishes and the 
positivity constraints is solely on $a_3$, 
\begin{align}
a_3\ge0 \Longrightarrow R^2 \le 1.
\end{align}
This is the disc section.
On the other hand, if the two $\mu_i$'s commute, the $d_{ijk}$ related term gives
$4x^2y^2$. The positivity constraint is on $a_4$
\begin{align}
a_4\ge0 \Longrightarrow R^4-2R^2-2x^2y^2+1\ge0.
\end{align}
This gives rise to the square section.

\section{volume estimation}
\label{sec:vol} 

The sets $M$ and $S$ are spheroids of varying radius and the set of
entangled states $M\backslash S$ is a spheroidal shell of varying thickness.
To get some idea of the volumes of $M$, $S$ and $M\backslash S$ and their
radial profiles we have performed Monte Carlo simulations on the sphere of
radius $\sqrt{3}$ in $\mathbb{R}^{15}$. The results are shown in Fig. \ref%
{fig:radial}, where we give the probabilities that a state $\vec{n}$ is
physical ({positive density matrix}) and that it is separable, as a function of $%
\left\vert \vec{n}\right\vert $, leading to a radial probability distribution function
for the physical states and the separable states. The distributions fall off
fairly abruptly, which allows us to identify a rough ``effective" radius of
about $\sqrt{3}/2$ for the physical states and about $0.45\times \sqrt{3}$
for the separable states.  Plotted in this way, the entangled states form a
rather thin shell. {This is not the only way to display the data, of course. 
If the volume densities were to be plotted instead, the function shown would 
be multiplied by $|\vec{n}|^{14}$, 
and the density of separable states would show a sharp peak. }

To estimate the volume of separable and physical states
inside the $15$ dimensional ball with radius $\sqrt{3}$, we first perform
simple sampling Monte Carlo to get the radial distribution function $p(r)$
at various $r$ values. Here $r=|\vec n|$. 
$10^{11}$ sample points are generated at each $r_k=
\sqrt{3}\times0.02 k$. The accuracy of each $p(r_k)$ is beyond five digits.
Uniformly distributed random numbers on a $14$ dimensional sphere is
generated using the algorithms in Ref. \cite{Watson83}.

Take $p(r)$ to be the radial distribution for the physical states for
example. To get $V_{\text{phys}}$ from $p(r)$, we perform numerical
integration using Simpson's rule. 
\begin{align}
V_{\text{phys}}\left[p(r)\right] =& \frac{\int_0^{\sqrt{3}}p(r) r^{14}dr}{%
\int_0^{\sqrt{3}} r^{14}dr}V_B \\
=& 15V_B\int_0^1 p(\tilde{r}) \tilde{r}^{14}d\tilde r. \label{eq:vol}
\end{align}
where $\tilde {r} = r/\sqrt{3}$ is the normalized radius, i.e., $|\vec{n}|/%
\sqrt{3}$. $V_B$ it the volume of a $15$-dimensional ball with radius $\sqrt{%
3}$

\begin{align}
V_B &\equiv V(r=\sqrt{3},s=15) \\
&= \frac{\pi^{s/2}}{\Gamma(s/2+1)}r^{s} \simeq 1444.905
\end{align}

From the radial probability distributions of Fig. \ref{fig:radial}, we get the volume of
separable states and physical states in the $15$-dimensional polarization
vector representation
\begin{align}
V_{\text{sep}}=& 0.008971\pm 0.000005 \\
V_{\text{phys}}=& 0.03700\pm 0.00001 \\
\frac{V_{\text{sep}}}{V_{\text{phys}}}=& 0.2424\pm 0.0002.
\end{align}%
 The separable states only account for about a quarter of the state space in
our volume measure, which is the Euclidean measure on $M$.  This measure has
the virtue that the distance between two different values of $\vec{n}$
corresponds to a physical distance between $15$ ``pointers" if a complete
measurement of all $15$ observables is made.  The volume in this measure is
different from that of Ref. \cite{Zyczkowski98}, where eigenvalues of the
density matrix were used to discriminate states.  With that approach the
separable states occupy more than half the volume of the full state space.

\begin{figure}[tbp]
\centering\includegraphics*[width=.8\linewidth]{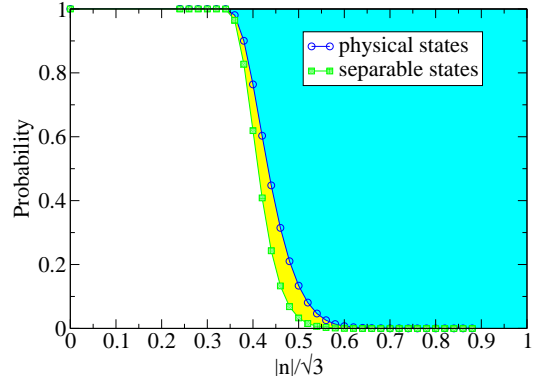}
\caption{Radial probability distribution functions of physical states 
(blue circle) and separable states
(green square) in the ball of radius $\protect\sqrt{3}$ in $\mathbb{R}^{15}$. 
For any fixed radius $|\vec{n}|$, the vertical span of the white, yellow and 
cyan regions denotes the probability of a randomly generated $\vec{n}$ being 
separable, entangled and unphysical. 
The volume of $S$ and $M$ can be calculated from Eq. \ref{eq:vol}.}
\label{fig:radial}
\end{figure}

A probability distribution function $P\left( C\right) $ for various $%
\left\vert \vec{n}\right\vert $ can be computed in a similar way. This is
shown in Fig. \ref{fig:diff_rad}. Again, appreciable entanglement is
concentrated at a fairly specific radius. At $|\vec{n}|=\sqrt{3}$, i.e., for
all pure states, the measure of separable states is zero. 
This is because pure separable states are parametrized by $4$ parameters while
pure states require $6$ parameters.
As stated above, 
$S$ contains a ball of radius $\left\vert \vec{n}\right\vert = 1/\sqrt{3}
\simeq0.57735$. 
The numerical study indicates that the maximum radius is at about 
$\left\vert \vec{n}\right\vert \simeq0.5782$. 
Although there is some angular variation in the function $C(\vec{n})$, 
the magnitude of $\vec n$ is a good rough indicator for the entanglement.

\begin{figure}[tbp]
\centering\includegraphics*[width=.8\linewidth]{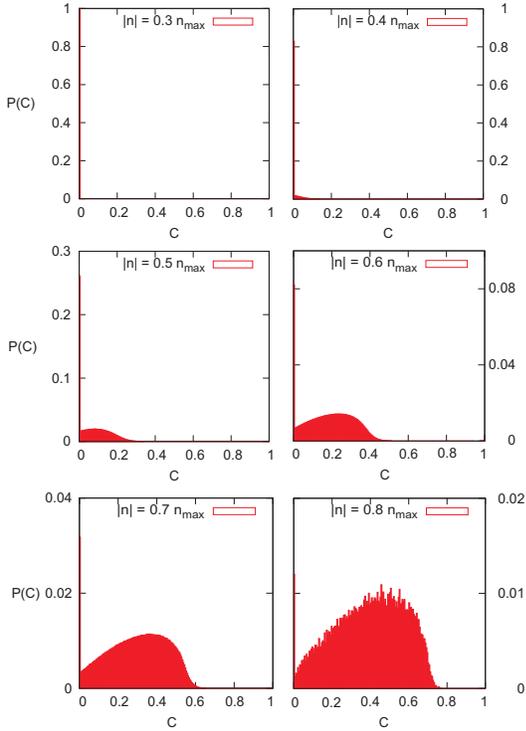}
\caption{
    Probability distribution of concurrence at different radius $|\vec n|$.  
For small $|\vec n|$, $P(C=0)$ is close to $1$, i.e., almost all states 
$\vec n$ are separable.
As $|\vec n|$ increases, more and more states are entangled and the probability
$P(C=0)$ keeps decreasing.
$n_{\max}=\protect\sqrt{3}$. }
\label{fig:diff_rad}
\end{figure}

\section{dynamical subspace $D$}
\label{sec:D} 

In the text we defined the dynamical subpace $D$ informally as the set of all 
trajectories allowed by the set of initial conditions and the equations
of motion.  Here we give a more formal definition.  

The authors of a model generally specify $n$ parameters in the total 
Hamiltonian for the coupled system and bath.  These parameters vary over some 
ranges, $g_{1,\min} \le g_1 \le  g_{1,\max}$, etc., giving a set $G=\{g_i\}$.
Furthermore, a set of initial conditions is also given by a similar set $H$.
For example, in the $YE$ model, $G =\{\Gamma\}$ with $0 \le \Gamma < \infty$, and 
$H =\{a_0\}$ with $0 \le a_0 \le 1$.  In the literature, $G$ and $H$ are almost 
always manifolds (or manifolds with boundaries), as they are in this example.  
Finally $T =\{t\}$ with $0 \le t < \infty$ is always a $1$-manifold.  
The equations of motion then define a smooth map $N: G \times H \times T 
\longrightarrow M$, where $M$ is the state space.  
The dynamical subspace $D$ is then defined as $D = \text{Image}(N)$.    

\begin{acknowledgments}
We thank Amrit De, Alex Lang and Andre L. Saraiva
for helpful discussions. This work was supported by
    Grant NSF-ECS-0524253 and NSF-DMR-0805045, by
    the DARPA QuEST program, and by ARO and LPS
    Grant W911NF-08-1-0482.

\end{acknowledgments}

\bibliography{/home/nos/projects/refs}

\end{document}